\documentclass[12pt,twoside]{article}
\usepackage[mathscr]{eucal}
\usepackage{amsmath,amsfonts,amssymb,amsthm}

\usepackage[matrix,arrow]{xy}
\usepackage{times}
\usepackage{graphicx}
\usepackage{epstopdf}
\usepackage{hyperref}

\advance\textheight\topskip
\renewcommand{\tilde}{\widetilde}
\renewcommand{\hat}{\widehat}
\newtheorem{prop}{Proposition}[section]

\newtheorem{theorem}[prop]{Theorem}

\renewcommand{\d}{\partial}

\newcommand{\RR}{\mathbb{R}}

\newcommand{\NN}{\mathbb{N}}

\def\cL{\mathcal{L}}

\def\cP{\mathcal{P}}
\def\cQ{\mathcal{Q}}
\def\cR{\mathcal{R}}

\numberwithin{equation}{section} \makeatletter
\@addtoreset{equation}{section}

\newcommand*\xbar[1]{%
  \hbox{%
    \vbox{%
      \hrule height 0.5pt 
      \kern0.3ex
      \hbox{%
        \kern-0.0em
        \ensuremath{#1}%
        \kern-0.0em
      }%
    }%
  }%
} 

\begin{document}

\def\mytitle{Poisson Structure of the Boundary Gravitons in 3D Gravity with Negative $\Lambda$.}

\pagestyle{myheadings} \markboth{\textsc{\small Troessaert}}{%
  \textsc{\small Canonical Structure of 3D Gravity}} \addtolength{\headsep}{4pt}

\begin{centering}

  \vspace{1cm}

  \textbf{\Large{\mytitle}}



  \vspace{1.5cm}

  {\large C\'edric Troessaert$^a$}

\vspace{.5cm}

\begin{minipage}{.9\textwidth}\small \it \begin{center}
   Centro de Estudios Cient\'ificos (CECs)\\ Arturo Prat 514,
   Valdivia, Chile \\ troessaert@cecs.cl \end{center}
\end{minipage}

\end{centering}

\vspace{1cm}

\begin{center}
  \begin{minipage}{.9\textwidth}
    \textsc{Abstract}. We use the hamiltonian formalism to study the asymptotic structure of 3 dimensional
    gravity with a negative cosmological constant. We start by defining very
    general fall-off conditions for the canonical variables and study the implied Poisson
    structure of the boundary gravitons. From the allowed differentiable gauge
    transformations, we can extract all the possible boundary conditions on
    the lagrange multipliers and the associated boundary hamiltonians. In the
    last section, we use this general framework to describe some of the
    previously known boundary conditions.
  \end{minipage}
\end{center}

\vfill

\noindent
\mbox{}
{\scriptsize$^a$Laurent Houart postdoctoral fellow.}

\thispagestyle{empty}
\newpage

\begin{small}
{\addtolength{\parskip}{-1.5pt}
 \tableofcontents}
\end{small}
\newpage

\section{Introduction}
\label{sec:introduction}

Since its introduction, Einstein's theory in three dimensions has been a very
useful toy model to study properties of gravitational theories. Even if it lacks some
features compared to its higher dimensional versions, like gravitational
waves, it still possesses dynamical objects \cite{Deser1984} and black-holes
\cite{Banados1992,Banados1993}. 

This theory is particularly interesting in the context
of AdS/CFT. In their seminal work \cite{Brown1986}, Brown and Henneaux showed that the algebra of
the conserved charges of asymptotically $AdS_3$ space-times is given by two
copies of the Virasoro algebra with non-zero central charge. This lead
to many interesting results, for instance:
Strominger was able to reproduce the Bekenstein-Hawking entropy of the BTZ
black-holes using the Cardy formula \cite{strominger1998}. Since then, this
framework has been extended: either by
relaxing the original asymptotic conditions of Brown-Henneaux
\cite{Porfyriadis2010,Troessaert2013} or introducing new asymptotics with
different boundary dynamics \cite{Compere2013,Avery2014}. We now have a few
different sets of boundary conditions available but it is reasonable to say
that a lot more possibilities should exist.

Using the Chern-Simons description of 3D gravity, one can solve the
constraints and obtain the reduced theory describing the dynamics of the
boundary gravitons. For Brown-Henneaux boundary conditions, this procedure
leads to a Liouville theory on the boundary
\cite{Coussaert1995,Henneaux2000,Rooman2001,Barnich2013}. On the other hand, for chiral
boundary conditions, one obtains a chiral Liouville theory on the boundary
\cite{Compere2013}. All these results rely
heavily on the fact that one can solve the constraints and are difficult
to generalize in different contexts.

\vspace{5mm}

In this work, we use the hamiltonian framework to provide a unified
description of the previously introduced boundary conditions. The idea is to
start with very general asymptotic fall-off conditions and use the results obtained in
\cite{Troessaert2013a}. In the process, we will build a description of the
reduced theory living on the boundary at infinity without explicitly solving
the constraints.

In the first section, we study the asymptotic structure of 3D gravity with a
negative cosmological constant. We introduce our asymptotic fall-off
conditions and
study the structure of the reduced phase-space. More precisely, we build
quantities parametrizing the boundary gravitons and compute the induced
poisson structure.

In the second section, we describe all possible boundary conditions on the
lagrange multipliers. These boundary conditions are responsible for the
dynamical part of the theory. In particular, they are in one to one
correspondence with the induced hamiltonian on the phase-space of the boundary gravitons.

In the last section, we use our formalism to describe some of the boundary
conditions previously obtained in the literature. We study both the conformaly
symmetric boundary conditions \cite{Troessaert2013,Brown1986} and the chiral
boundary conditions \cite{Avery2014,Compere2013}

\vspace{5mm}

In \cite{Apolo2014}, the authors conjectured that all the previously
introduced asymptotic conditions for 3D gravity are dual to Polyakov 2D gravity
with different gauge choices for the metric. It would be interesting to see how
their approach can be extended to the most general asymptotic conditions introduced
here.

\vspace{5mm}

In this paper, we use the notation $O(r^n)$ to describe functions with the
following behavior in the limit $r\rightarrow \infty$:
\begin{equation}
	f(r,x^A) = O(r^n) \quad \Rightarrow \quad \lim_{r\rightarrow \infty}
	\frac{f}{r^n} = \bar f(x^A).
\end{equation}
We will also ask for a compatible behavior with as many partial derivatives as
needed:
\begin{gather}
	f(r,x^A) = O(r^n) \quad \Rightarrow \quad \d^k_r f(r,x^A) = O(r^{n-k})\quad \text{and} \quad \d^k_A f = O(r^n).
\end{gather}

\section{Asymptotic structure}
\label{sec:AsymptoticStructure}

The bulk hamiltonian action for gravity in 3 dimension is given by:
\begin{eqnarray}
\label{eq:bulkaction}
S[N,N^i,g_{ij},\pi^{ij}] & = & \frac{1}{16\pi G}\int dt \int_\Sigma d^2x\, \left\{\pi^{ij} \d_t
g_{ij} - N \cR - N^i \cR_i\right\},\\
\cR & =& - \sqrt{g} \left[ R - 2\Lambda  + \frac{1}{g} \left(
    \pi^2 - \pi^{ij} \pi_{ij} \right)\right], \\
\cR_i & = & -2 \nabla_j\pi^{\phantom{i}j}_i,
\end{eqnarray}
where $g_{ij}$ is a 2 dimensional metric and $\pi^{ij}$ is a
density. In order to apply the formalism of \cite{Troessaert2013a},
we need boundary conditions on the dynamical variables
$(g_{ij},\pi^{ij})$. As, we want to study the
asymptotic structure, we need fall-off conditions in order to have
generators given by finite quantities. The most common choice is the
one used in \cite{Brown1986} but there have been other propositions
\cite{Porfyriadis2010,Compere2013,Troessaert2013,Avery2014}. Following the
results of \cite{Troessaert2013a}, we expect these boundary
conditions to share the same reduced phase-space, the differences
being in the choice of the Hamiltonian.

We will start with general fall-off conditions on the phase-space
containing all of the previously proposed boundary conditions. The
analysis of the boundary conditions on the lagrange multipliers will
be posponed to the study of the hamiltonian generators starting in
section \ref{sec:conf}. We will consider the following asymptotic behavior:
\begin{gather}
\label{eq:assympg}
g_{rr} = \frac{l^2}{r^2} + O(r^{-4}), \quad g_{r\phi} = O(r^{-1}),
\quad g_{\phi\phi} = r^2 \bar \gamma(t, \phi) + O(1),\\
\pi^{rr} = O(r),\quad \pi^{r\phi} = O(r^{-2}), \quad \pi^{\phi\phi} = O(r^{-5}), 
\label{eq:assympp}
\end{gather}
where $\Lambda = -\frac{1}{l^2}$ and $\bar \gamma$ is a dynamical
field which is always positive. In \cite{Brown1986}, the authors
showed that such fall-off conditions are not enough for a hamiltonian
analysis of the problem. We also have to impose the constraints asymptotically:
\begin{equation}
\label{eq:assympR}
\cR = O(r^{-n}), \quad \cR_i = O(r^{-n}) \qquad \forall n\in \NN.
\end{equation}
With this set of fall-off conditions, the bulk part of the action
\eqref{eq:bulkaction} is finite whenever the lagrange multipliers
satisfy
\begin{equation}
\exists m \in \NN\quad \text{s.t.}\quad  N = O(r^m), \quad N^i = O(r^m).
\end{equation}

The additional conditions on the constraints \eqref{eq:assympR} have
some useful consequences. In particular, we have:
\begin{equation}
\label{eq:asymppirr}
\pi^{rr} =  \frac{r}{2 l} P(t, \phi) + O(r^{-1}),
\end{equation}
and
\begin{equation}
\label{eq:asympK}
\d_r \left(r^2 (K + \frac{1}{l}) \right) = O(r^{-3}),
\end{equation}
where $K$ is the trace of the extrinsic curvature of the circles $r$
equals constant (see appendix \ref{sec:ADMspa}).

\subsection{Differentiable gauge transformations}
\label{sec:inf-diff-gauge-gener}

Gauge-like transformations are given by:
\begin{equation}
g_{ij} = \frac{\delta ( \xi\cR + \xi^k\cR_k)}{\delta
  \pi^{ij}}, \quad \pi^{ij} =-
\frac{\delta (\xi \cR + \xi^k \cR_k)}{\delta g_{ij}},
\end{equation}
where the gauge parameters $\xi, \xi^i$ can depend on the fields. We
will restrict our analysis to gauge parameters with the following
asymptotic behavior:
\begin{equation}
\label{eq:inf-paramgauge}
\xi = O(r), \qquad \xi^r = O(r), \qquad \xi^\phi = O(1).
\end{equation}
In this case, using \eqref{eq:assympR}, the explicit form of the
gauge-like transformations is worked out to be:
\begin{eqnarray}
\label{eq:gaugeg}
\delta_\xi \pi^{ij} & = & -\sqrt{g} \xi
\Lambda g^{ij}  + \sqrt{g} \left(
  \nabla^i\nabla^j \xi - g^{ij} \nabla^k\nabla_k\xi\right)\nonumber\\
&& - 2\frac{\xi}{\sqrt{g}}\left( \pi^{ik}\pi_k^j - 
  \pi\pi^{ij} \right) -
\frac{\xi}{\sqrt{g}}\frac{g^{ij}}{2}\left(\pi^2 -
  \pi^{kl}\pi_{kl} \right),\nonumber\\
&& +\d_k \left( \xi^k \pi^{ij}\right) - \d_k\xi^i \pi^{kj} - \d_k\xi^j \pi^{ki}+ O(r^{-n}),\\
\label{eq:gaugep}
\delta_\xi g_{ij} & = & 2
\frac{\xi}{\sqrt{g}} \left( \pi_{ij} - \pi g_{ij}\right) +
\xi^k \d_k g_{ij} +
\d_i \xi^k g_{kj} + \d_j \xi^k g_{ki} + O(r^{-n}),
\end{eqnarray}
for all $n \in \RR$.

\vspace{5mm}

A differentiable gauge transformation is a gauge-like transformation $\delta_\xi$
for which we can associate a differentiable generator. This requires two
conditions to be met: the transformation $\delta_\xi$ preserves the boundary conditions
and the generator $\Gamma_\xi$ satisfies
\begin{equation}
\delta \Gamma_\xi = \int_\Sigma d^2x \left(\frac{\delta
    \Gamma_\xi}{\delta g_{ij}} \delta g_{ij} +  \frac{\delta
    \Gamma_\xi}{\delta \pi^{ij}} \delta \pi^{ij}\right).
\end{equation}
To compute the set of differentiable gauge transformations, we will
start by computing the set of gauge-like transformations
\eqref{eq:gaugeg}-\eqref{eq:gaugep} preserving the boundary conditions.

The variations of the constraints under a gauge-like transformation are
given by:
\begin{eqnarray}
\delta_\xi \cR = \d_i (\xi^i \cR) - \d_i (\xi g^{ij}\cR_j) - \d_i \xi
g^{ij} \cR_j + O(r^{-n}), \qquad \forall n \in \RR,\\
\delta _\xi \cR_k = \d_k \xi \cR + \d_i(\xi^i \cR_k) + \d_k \xi^i
\cR_i+ O(r^{-n}),\qquad \forall n \in \RR.
\end{eqnarray}
We see that any transformation of the form \eqref{eq:inf-paramgauge} will
preserve the fall-off conditions on the constraints. Computing the
variation of the metric and using the fall-off conditions, we obtain:
\begin{eqnarray}
\delta_\xi g_{rr} & = & 2 \frac{l^2}{r^2}(\d_r \xi^r-\frac{1}{r}\xi^r )
+ 2 \d_r \xi^\phi g_{r\phi} + O(r^{-4}),\\
\delta_\xi g_{r\phi} & = & r^2 \sigma \d_r \xi^\phi + O(r^{-1}),\\
\delta_\xi g_{\phi\phi} & = & -2 \xi \sqrt{g} \pi^{rr} + \xi^r
\d_rg_{\phi\phi} + \xi^\phi \d_\phi g_{\phi\phi} + 2 \d_\phi \xi^\phi
g_{\phi\phi} + O(1).
\end{eqnarray}
The preservation of the fall-off conditions for $g_{rr}$ and
$g_{r\phi}$ leads to
\begin{equation}
\xi^r = r \psi + O(r^{-1}), \quad \xi^\phi = Y + O(r^{-2}),
\end{equation}
where $\psi$ and $Y$ are arbitrary functions independant of $r$.
Taking this into account
and using the spatial 1+1 decomposition of the metric described in 
appendix \ref{sec:ADMspa}, the variations of the momenta become
\begin{eqnarray}
\delta_\xi \pi^{rr} & = & - \sqrt{\sigma}
  \frac{r^3}{l^3}(\d_r \xi - \frac{1}{r} \xi)+ O(r),\\
\delta_\xi \pi^{r\phi} & = & \frac{1}{l\sqrt{\sigma}} (\d_\phi+\frac{r}{l^2} \lambda_\phi )(\d_r \xi-\frac{1}{r} \xi)+ O(r^{-2}),\\
\delta_\xi \pi^{\phi\phi} &=& - \frac{1}{\sqrt{g}} \left( \d_r^2 \xi - \frac{\lambda^2
    }{l^2} \xi  - \frac{\d_r \lambda}{\lambda}\d_r
  \xi + \frac{r}{l^2} \lambda_\phi \lambda^\phi (\d_r \xi-\frac{1}{r} \xi)\right. \nonumber \\ && \qquad
\left. -(3\frac{\lambda^\phi}{r} +\d_r \lambda^\phi) \d_\phi
  \xi\right) + O(r^{-5}) \label{eq:varpiphiphiI}
\end{eqnarray}
where $\lambda_\phi = O(r^{-1})$ and $\lambda = \frac{l}{r} + O(r^{-3})$. The preservation of the fall-off
conditions for $\pi^{rr}$ and $\pi^{r\phi}$ then imply
\begin{equation}
\xi = r f + \kappa,\qquad \kappa = O(r^{-1}),
\end{equation}
where $f$ is another arbitrary function independent of $r$. Using this
and the asymptotic form of $\pi^{rr}$ given in \eqref{eq:asymppirr},
the variation of $g_{\phi\phi}$ automatically preserves the fall-off
condition \eqref{eq:assympg}.

The only condition we still need to check is the preservation of
$\pi^{\phi\phi}=O(r^{-5})$.
Using the expansion $\lambda = \frac{l}{r} + \tilde \lambda$
with $\tilde \lambda = O(r^{-3})$, we can simplify the variation \eqref{eq:varpiphiphiI} to:
\begin{equation}
\label{eq:varfinalpiphiphi}
\delta_\xi \pi^{\phi\phi} 
= - \frac{r^{-2}}{\sqrt{g}} \d_r\left[r^2\left( \d_r\kappa -
    \frac{1}{r}\kappa - \frac{\tilde \lambda}{l} rf -\lambda^\phi
    r\d_\phi f\right)\right] + O(r^{-5}). 
\end{equation}
Any possible transformation satisfying this condition will also
preserve the more constrained form of $\pi^{rr}$ given in equation
\eqref{eq:asymppirr}. Computing the variation of $\pi^{rr}$ taking
the gap into account leads to:
\begin{equation}
\label{eq:varfinalpirr}
\delta_\xi \pi^{rr} = \frac{\gamma}{\lambda l}\left( \d_r \kappa
  -\frac{1}{r} \kappa - \frac{\tilde \lambda}{l} r f - \lambda^\phi r\d_\phi f - l( K+
  \frac{1}{l}) f\right) +  \omega r + O(r^{-1}),
\end{equation} 
where $\omega$ is a function independent of $r$ encoding part
of the variation of $P$. In order to preserve the asymptotic form
of $\pi^{rr}$, the function $\kappa$ must be of the form
\begin{eqnarray}
\kappa &=& - \frac{l^2}{2r} \chi - r\int_r^\infty dr' j(r') + O(r^{-3}), \\
j &=& \frac{\tilde \lambda}{l} f + \lambda^\phi\d_\phi f + \frac{l}{r}( K+
  \frac{1}{l}) f = O(r^{-3}),\label{eq:fuckingj}
\end{eqnarray}
where $\chi$ is an arbitrary function independent of $r$. Combining
equation \eqref{eq:varfinalpirr} with \eqref{eq:asympK}, we see that
such a $\kappa$ induces a variation \eqref{eq:varfinalpiphiphi} that
automatically preserves the fall-off of $\pi^{\phi\phi}$. 
 We have shown the following:
\begin{theorem}
The set of gauge-like transformations preserving the asymptotic
conditions \eqref{eq:assympg}-\eqref{eq:assympR} is given by:
\begin{eqnarray}
\label{eq:allowed}
\xi & = & r f - \frac{l^2}{2r} \chi - r\int_r^\infty dr' j(r') +
O(r^{-3}),\\
\xi^r & = & r \psi + O(r^{-1}),\\
\xi^\phi & = & Y + O(r^{-2}),
\end{eqnarray}
where the function $j$ is given in equation \eqref{eq:fuckingj} and the
four functions $f, \chi, \psi$ and $Y$ are independent of $r$.
\end{theorem}

The second condition for a gauge-like transformation to be differentiable is
the existance of a differentiable generator.
The bulk part of the generator of a gauge-like transformation is given
by the smeared constraints:
\begin{equation}
\tilde \Gamma_{\xi} = \frac{1}{16\pi G}\int_\Sigma d^2x \left( \xi \cR + \xi^i \cR_i\right).
\end{equation}
The boundary term coming from a general variation is then easily
computed:
\begin{eqnarray}
\label{eq:nonintegI}
\delta \tilde \Gamma_\xi &=& \int_\Sigma d^2x \, \left( \frac{\delta \tilde
    \Gamma_\xi}{\delta g_{ij}} \delta g_{ij} + \frac{\delta \tilde
    \Gamma_\xi }{\delta \pi^{ij}} \delta \pi^{ij}
\right)\nonumber \\
&& + \frac{1}{16\pi G}\oint_{\d \Sigma} (d^1x)_k \, \left\{-2 \xi^i \delta \pi^k_{\phantom k i}
  +\xi^k \pi^{ij} \delta g_{ij} - \xi \sqrt{g} \left(g^{ij}\delta \Gamma^k_{ij} -
    g^{ki}\delta \Gamma^j_{ji}\right)\right. \nonumber\\
&& \qquad \left. + \d_l \xi \sqrt{g}
  \left(g^{ki}g^{lj} - g^{kl} g^{ij} \right) \delta  g_{ij} + \Theta^k(\cR, \cR_i)\right\}.
\end{eqnarray}
The function $\Theta$ is coming from the variation of the gauge
parameters $\xi, \xi^i$: it is a local function of the constraints and
their derivatives. In this case, as we have imposed the
constraints asymptotically, it will always be zero. Inserting our
fall-off conditions, the asymptotic form of the gauge parameters and
evaluating at the boundary $r\rightarrow \infty$,
the boundary term becomes: 
\begin{multline}
\label{eq:nonintegII} 
-\frac{1}{16\pi G} \oint_{\d \Sigma} d\phi\, \lim_{r \rightarrow \infty}\left\{2Y \delta
  \pi^r_\phi +l \psi \, \delta P +2 r^2 f \delta\left( \sqrt{\bar\gamma} (K +
  \frac{1}{l})\right) \right.  \\ \left. + 2 \left(
 r^2 (K + \frac{1}{l})f +l\,  \chi\right) \delta
    \sqrt{\bar\gamma}\right\}.
\end{multline}
Let's introduce the fields:
\begin{equation}
J(t, \phi) \equiv \frac{2}{l}\lim_{r\rightarrow \infty}  \pi^r_\phi,
\qquad  M(t, \phi) \equiv
 \frac{ 2 \sqrt{\bar\gamma}}{l} \lim_{r\rightarrow \infty} \left(r^2 ( K +
  \frac{1}{l})\right), \qquad Q(t, \phi) \equiv 2 \sqrt{\bar\gamma}.
\end{equation}
The boundary term \eqref{eq:nonintegII} is integrable if and only if there
exists a functional on the circle
\begin{equation}
\frac{l}{16\pi G} \oint_{\d \Sigma}
d\phi\, k_\xi( P, J, M, Q),
\end{equation}
such that:
\begin{gather}
Y  =  \frac{\bar \delta k_\xi}{\delta J},\qquad
\psi  =  \frac{\bar \delta k_\xi}{\delta  P},\label{eq:asymparamkI}\\
f  =  \frac{\bar \delta k_\xi}{\delta  M},\qquad
\tilde \chi \equiv \chi + \frac{ M}{ Q} f  =  \frac{\bar \delta k_\xi}{\delta  Q},
\label{eq:asymparamkII}
\end{gather}
where the Euler-Lagrange derivative $\frac{\bar \delta}{\delta}$ is the one
defined on the circle only:
\begin{equation}
\frac{\bar \delta k}{\delta M} = \sum_k (-\d_\phi)^k
\frac{\d k}{\d \d^k_\phi M}.
\end{equation}
If such a functional exists, the differentiable generator of the transformation is given by:
\begin{equation}
\Gamma_\xi = \frac{1}{16\pi G}\int_\Sigma d^2x \left( \xi \cR + \xi^i \cR_i\right) + \frac{l}{16\pi G} \oint_{\d \Sigma}
d\phi\, k_\xi(P,  J, M, Q).
\end{equation}
On the constraints, we obtain 
\begin{equation}
\Gamma_\xi \approx \frac{l}{16\pi G} \oint_{\d \Sigma}
d\phi\, k_\xi(P, J, M, Q).
\end{equation}
The transformations for which $\Gamma_\xi \approx 0$ are called proper
gauge transformations. They are the true gauge freedom of the system
as they are generated by constraints and always comute with the
differentiable Hamiltonian \cite{Troessaert2013a}. In the following,
we will denote the parameters of proper gauge transformations by
$\eta$ and $\eta^i$.

The set differentiable gauge transformations form an algebra under the
Poisson bracket for which the set of proper gauge transformations is
an ideal. We have proved that
\begin{theorem}
The quotient of the differentiable gauge transformation by the proper
gauge transformations is parametrized by the functionals of $P,
 J, M$ and $Q$ defined on the circle:
\begin{equation}
\frac{l}{16\pi G} \oint_{\d \Sigma} d\phi\, k_\xi(P, J, M, Q).
\end{equation}
\end{theorem}
\noindent The induced Poisson bracket on the quotient will be computed
in section \ref{sec:dirac-brack-bound}.

\subsection{Boundary gravitons}
\label{sec:boundary-gravitons}

We expect the quantities $P, J, M$ and $Q$ that we defined
in the previous section to encode all the information about the
boundary gravitons. More specifically, we expect them to be gauge
invariant and to completely characterize the configuration up to proper
gauge transformations. 

The parameters of proper gauge transformations $\Gamma_\eta$ have the following fall-off
\begin{equation}
\eta = O(r^{-3}), \qquad \eta^r = O(r^{-1}), \qquad \eta^\phi = O(r^{-2}).
\end{equation}
We easily show that the associated transformations on the relevant canonical
fields are given by:
\begin{equation}
\delta_\eta \pi^{rr} = O(r^{-1}), \qquad \delta_\eta \pi^{r\phi} =
O(r^{-4}), \qquad \delta_\eta g_{\phi\phi} = O(1).
\end{equation}
This means that $P$, $J$ and $Q$ are gauge invariant quantities. For
$M$, we need the transformation law of $K$ (see eq \eqref{eq:transfoK}). A straightforward computation gives
\begin{equation}
\delta_\eta K = O(r^{-4}),
\end{equation}
which means that $M$ is also gauge invariant.

In order to analyse the structure of the reduced phase-space, it is
easier to fix the gauge. The simplest choice is the Fefferman-Graham
gauge which is given by:
\begin{equation}
g_{rr} = \frac{l^2}{r^2}, \qquad g_{r\phi} = 0, \qquad \pi^{\phi\phi}
= 0.
\end{equation}
This gauge can always reached by a proper gauge transformation (more details
are given in appendix \ref{sec:FGfixing}). 
With the gauge fixed, the constraints simplify
drastically: 
\begin{eqnarray}
\label{eq:gaugefixconstrr}
\cR_r & = & -2 \frac{l^2}{r^2} \left( \d_r \pi^{rr} - \frac{1}{r}
  \pi^{rr} + \d_\phi \pi^{r\phi} \right),\\
\label{eq:gaugefixconstrphi}
\cR_\phi & = & -2 \gamma\frac{l}{r}\left( \frac{r}{l}\d_r \pi^{r\phi} + \frac{2}{l}
  \pi^{r\phi} - 2\pi^{r\phi} (K + \frac{1}{l})\right),\\
\label{eq:gaugefixconstrperp}
\cR & = & -2 \frac{l}{r}\sqrt{\gamma}\left( \frac{r}{l} \d_r K - K^2 -
(\pi^{r\phi})^2 + \frac{1}{l^2}\right),
\end{eqnarray}
where the extrinsic curvature is given by $K = -\frac{r}{2l}
\gamma^{-1} \d_r \gamma$ (see appendix \ref{sec:ADMspa}). This gives us a set of four differential
equation in $r$ for which $P, J, M$ and $Q$ are the
corresponding four integration constants. This can be seen easily
as this system is solvable explicitly. 

In term of $L_\pm = K +
\frac{1}{l} \pm \pi^{r\phi}$, we can rewrite the constraints
\eqref{eq:gaugefixconstrphi} and \eqref{eq:gaugefixconstrperp} as
\begin{equation}
\frac{r}{l}\d_r L_\pm + \frac{2}{l} L_\pm - L_\pm^2=0.
\end{equation}
This gives
\begin{equation}
L_{\pm}= \frac{2}{l}\frac{ A_\pm}{ A_\pm + \frac{r^2}{l^2}} = 2l\frac{A_\pm}{r^2}+O(r^{-4}),
\end{equation}
where $A^\pm$ are two integration constants.
We can then solve
for $\pi^{rr}$ and $\gamma$:
\begin{eqnarray}
\pi^{rr} &=& r \frac{P}{2l} + \frac{r}{2l}\left(\frac{\d_\phi
    A_+}{ A_+
    + \frac{r^2}{l^2}} -\frac{\d_\phi A_-}{A_-
    + \frac{r^2}{l^2}} \right) = r \frac{P}{2l}+O(r^{-1}),\\
\gamma & = & \bar \gamma r^{2}\left(1 + \frac{l^2}{r^2}A_+ \right)\left(1 +
  \frac{l^2}{r^2}A_- \right) = \bar\gamma r^{2} + O(1),
\end{eqnarray}
with the last two integration constants $\bar\gamma = \frac{Q^2}{4}$ and
$P$. The functions $A_\pm$ are related to
$M$ and $J$ by:
\begin{equation}
J = 2 \bar\gamma (A_+ - A_-), \qquad M = 2 \sqrt{\bar\gamma}(A_+ + A_-).
\end{equation}
\begin{theorem}
The four functions $P, J, M$ and $Q>0$ completely determine
the configuration asymptotically up to gauge transformations. They parametrize the only degrees of freedom of the
theory: the boundary gravitons.
\end{theorem}
\noindent The above analysis was only done asymptotically. For specific values of $P, J, M$ and $Q$, we have no guaranty
that the configuration will be regular everywhere in the bulk.

The BTZ black-holes \cite{Banados1992,Banados1993} are given by:
\begin{equation}
\label{eq:BTZinf}
P=0, \qquad J = 8 G \,\frac{j}{l}, \qquad M =  8 G \,m,\qquad Q = 2,
\end{equation}
where $m$ and $j$ are the mass and angular momentum of
black-hole. Let's remark that we are only talking about a
configuration at fixed $t$. To have the full 3D black-hole, we also
need the right time evolution: the right Hamiltonian. This will be
studied in section \ref{sec:brown-henn-bound}.

\subsection{Dirac bracket for the boundary gravitons}
\label{sec:dirac-brack-bound}
The Poisson bracket of two differentiable functionals
$F[g_{ij},\pi^{ij}]$ and $G[g_{ij},\pi^{ij}]$ is given by
\begin{equation}
\left\{F[g_{ij},\pi^{ij}], G[g_{ij},\pi^{ij}] \right\} = 16 \pi G\int_{\Sigma}
d^2x \, \left( \frac{\delta F}{\delta g_{ij}}\frac{\delta G}{\delta \pi^{ij}} -\frac{\delta G}{\delta g_{ij}}\frac{\delta F}{\delta \pi^{ij}} \right).
\end{equation}
For differentiable gauge generators, a straightforward computation gives
\begin{eqnarray}
\label{eq:constalge}
\left\{ \Gamma_\xi, \Gamma_\zeta\right\} & = & \tilde G \left[ [\xi,\zeta]_g\right]\\ && +\frac{1}{16\pi G}
\oint_{\d \Sigma} (d^{n-1}x)_k \left\{ 2(\zeta^k \nabla_i\xi_j
  -\xi^k\nabla_i\zeta_j) \pi^{ij}+ 2 \left[ \xi, \zeta \right]^j_{SD} \pi^k_j\right. \nonumber\\ &&\quad  \left. +2\sqrt{g}\left(
 \nabla_i \xi^k \nabla^i\zeta - \nabla_i \xi^i \nabla^k\zeta -
 \nabla_i \zeta^k \nabla^i\xi + \nabla_i \zeta^i \nabla^k\xi\right)
\right.\nonumber\\ &&\quad  \left. -
  (\zeta\xi^k-\xi\zeta^k) (2 \Lambda \sqrt{g} -
  \frac{1}{\sqrt{g}}(\pi^2 - \pi^{ij}\pi_{ij})) + \Theta^k(\cR, \cR_i)\right\},\nonumber\\
\left[ \xi, \zeta \right]^a_{g} & = & \left[ \xi, \zeta \right]^a_{SD} +
\delta_\zeta \xi^a - \delta_\xi \zeta^a + \Xi^a(\cR, \cR_i),
\end{eqnarray}
where $\xi^a=(\xi, \xi^i)$ and the functions $\Theta$ and $\Xi$ are
local functions of the contraints and their derivatives. The surface deformation bracket is given by:
\begin{eqnarray}
\left[ \xi, \zeta \right]_{SD}&=&\xi^i\d_i \zeta - \zeta^i \d_i \xi,\\
\left[ \xi, \zeta \right]^i_{SD}& = & \xi^j\d_j \zeta^i - \zeta^j\d_j \xi^i + g^{ij}
 \left( \xi \d_j \zeta - \zeta \d_j \xi\right).
\end{eqnarray}

Differentiable gauge generators are first-class functionals,
evaluating their Poisson bracket will also give us their Dirac bracket
when evaluated on the reduced phase-space. Let's consider two
differentiable gauge generators $\Gamma_1$ and $\Gamma_2$ associated
to the functionals
\begin{equation}
\frac{l}{16\pi G} \oint_{\d \Sigma}d\phi\, k_1(P, J, M, Q) \qquad \text{and} \qquad \frac{l}{16\pi G} \oint_{\d \Sigma}d\phi\, k_2(P, J, M, Q).
\end{equation}
The corresponding gauge parameters $\xi_1$ and $\xi_2$ are given, up to
proper gauge transformations, by the identifications \eqref{eq:asymparamkI} and
\eqref{eq:asymparamkII}. By construction, we then have the following
\begin{equation}
\left\{ \frac{l}{16\pi G} \oint_{\d \Sigma}d\phi\, k_1(P, J, M, Q), \frac{l}{16\pi G} \oint_{\d
    \Sigma}d\phi\, k_2(P, J, M, Q)\right\}^* \approx \left\{ \Gamma_1,
\Gamma_2\right\},
\end{equation}
where the LHS is the bracket on the reduced phase space. On the
constraints surface, the RHS reduces to the boundary term of
\eqref{eq:constalge}. It is a gauge invariant quantity, it is easier
to evaluate it when the gauge is fixed. Using the Fefferman-Graham
gauge described in the previous section, we obtain
\begin{eqnarray}
\left\{ \Gamma_1, \Gamma_2\right\} & \approx & \frac{l}{16\pi G} \oint_{\d \Sigma} d\phi \,
\left\{P (Y_1 \d_\phi \psi_2 +f_1 \tilde \chi_2) + JY_1 \d_\phi Y_2 +
  \frac{4 J}{Q^2} f_1 \d_\phi f_2 \right. \nonumber \\ && \qquad  +M (Y_1 \d_\phi f_2 + \psi_1 f_2) + Q (Y_1 \d_\phi \tilde
  \chi_2 - \psi_1 \tilde \chi_2)  \nonumber \\ && \qquad \left. + \frac{4}{Q} \d_\phi \psi_1 \d_\phi
  f_2 - (1 \leftrightarrow 2) \right\}.
\end{eqnarray}
If we replace, $Y, f, \psi$ and $\tilde \chi$ by their values in term
of the Euler-Lagrange derivatives of $k_1$ and $k_2$ using
\eqref{eq:asymparamkI} and \eqref{eq:asymparamkII}, we obtain the
induced Dirac bracket as
\begin{multline}
\label{eq:asympboundbracket}
\left\{ \Gamma_1, \Gamma_2\right\}  \approx  \frac{l}{16\pi G} \oint_{\d \Sigma} d\phi \,
\left\{P \left(\frac{\bar \delta k_1}{\delta J} \d_\phi \frac{\bar \delta k_2}{\delta P} +\frac{\bar \delta k_1}{\delta M} \frac{\bar \delta k_2}{\delta Q}\right)
\right. \\ 
  + M \left (\frac{\bar \delta k_1}{\delta J} \d_\phi \frac{\bar \delta k_2}{\delta M} + \frac{\bar \delta k_1}{\delta P}\frac{\bar \delta k_2}{\delta M}\right) 
  + Q \left(\frac{\bar \delta k_1}{\delta J} \d_\phi \frac{\bar \delta k_2}{\delta Q} - \frac{\bar \delta k_1}{\delta P} \frac{\bar \delta k_2}{\delta Q}\right) \\
  \left. + J\left( \frac{\bar \delta k_1}{\delta J}\d_\phi \frac{\bar \delta k_2}{\delta J} +
  \frac{4}{Q^2}\frac{\bar \delta k_1}{\delta M} \d_\phi \frac{\bar \delta
  k_2}{\delta M}\right)+ \frac{4}{Q} \d_\phi \frac{\bar \delta k_1}{\delta P} \d_\phi
  \frac{\bar \delta k_2}{\delta M} - (1 \leftrightarrow 2) \right\}.
\end{multline}

\section{Boundary hamiltonian}
\label{sec:boundcondlagr}

As shown in \cite{Troessaert2013a}, the differentiable Hamiltonian is given by the
boundary conditions on the Lagrange multipliers. More precisely, the
Hamiltonian for 3D gravity is given by the differentiable gauge
generator associated to the gauge parameters $N$ and $N^i$. We saw in
section \ref{sec:inf-diff-gauge-gener} that, on the constraints surface, it is given by a boundary term
\begin{equation}
H[g_{ij}, \pi^{ij}] \approx \frac{l}{16\pi G} \oint_{\d \Sigma} k_H(M, J, P, Q),
\end{equation}
with
\begin{gather}
f_H \equiv \lim_{r\rightarrow \infty} \frac{N}{r} = \frac{\bar \delta
  k_H}{\delta M}, \qquad \psi_H \equiv \lim_{r\rightarrow
  \infty} \frac{N^r}{r} = \frac{\bar \delta k_H}{\delta P},\\
Y_H \equiv \lim_{r\rightarrow \infty} N^\phi= \frac{\bar \delta
  k_H}{\delta J}, \qquad \tilde\chi_H \equiv \lim_{r\rightarrow
  \infty} \frac{r}{l} \left( \frac{1}{\lambda} \d_r N - \frac{1}{l} N
  - \frac{r}{l}\lambda^\phi \d_\phi N\right)= \frac{\bar \delta
  k_H}{\delta Q}.
\end{gather}

Tuning these boundary conditions we can build any
functional $k_H$ on the boundary. This is our main result:
\begin{theorem}
If we assume that the canonical variables have the following asymptotic
behavior:
\begin{gather}
	g_{rr}=\frac{l^2}{r^2} + O(r^{-4}), \quad g_{r\phi}=O(r^{-1}),
	\quad g_{\phi\phi}=r^2\bar\gamma(t,\phi)+O(1),\\
	\pi^{rr} = O(r), \quad\pi^{r\phi} = O(r^{-2}), \quad \pi^{\phi\phi}
	=O(r^{-5},\\
	\cR = O(r^{-n}),\quad \cR_i = O(r^{-n})\qquad \forall n \in \RR.
\end{gather}
then the set of possible boundary conditions at spatial infinity on the lagrange multipliers $(N,
N^i)$ is in one to one correspondance with the functionals
$\oint_{\d\Sigma}k_H(M, J, P, Q)$ (modulo the constant
functionals) where the boundary 
fields are defined by:
\begin{eqnarray}
	P(t, \phi) \equiv 2l \lim_{r\rightarrow \infty} \frac{
	\pi^{rr}}{r}, \quad
J(t, \phi) \equiv \frac{2}{l}\lim_{r\rightarrow \infty}  \pi^r_\phi,
\\  M(t, \phi) \equiv
 \frac{ 2 \sqrt{\bar\gamma}}{l} \lim_{r\rightarrow \infty} \left(r^2 ( K +
  \frac{1}{l})\right), \qquad Q(t, \phi) \equiv 2 \sqrt{\bar\gamma},
\end{eqnarray}
wich $\bar \gamma >0$.
On the constraint's surface, we obtain a theory on the boundary $\d\Sigma$
with a phase-space parametrized by $(M, J, P, Q)$ with a
bracket given in equation \eqref{eq:asympboundbracket} and an Hamiltonian given by
\begin{equation}
H[g_{ij}, \pi^{ij}] \approx \frac{l}{16\pi G} \oint_{\d \Sigma} k_H(M, J, P, Q),
\end{equation}
\end{theorem}
\noindent This analysis only concerns the differentiable structure at infinity, we
didn't treat any of the possible obstruction coming from the bulk structure of
the space-time.

A surprising feature is the need for 4 functions in order to completely
describe the asymptotic phase-space. When written in term of Chern-Simons
theory, one needs 6 functions to describe the corresponding asymptotic phase-space. Since
one adds 3 gauge degrees of freedom in the bulk, one would have expected to have
three more asymptotic functions in the Chern-Simons description compared to
the metric description. 

\vspace{5mm}

We will now study the different type of boundary conditions
that appeared in the literature. We will start with the sets of boundary
conditions that have the conformal algebra in two dimensions as a
symmetry algebra. 

\section{Some examples of boundary conditions}

\subsection{Conformal}
\label{sec:conf}

Let's consider the boundary conditions presented in
\cite{Troessaert2013}. With the coordinates $x^A=t, \phi$, they are given by:
\begin{eqnarray}
\label{eq:bcconfextI}
g_{rr} & = & \frac{l^2}{r^2} + C_{rr}r^{-4} + o(r^{-4}),\\
g_{rA} & = & C_{rA}r^{-3} + o(r^{-3}),\\
g_{AB} & = & r^2 e^{2\varphi}\eta_{AB} + C_{AB} + o(1),\\
\label{eq:confaddcond}
0 & = & e^{-2\varphi}\eta^{AB}C_{AB} + \frac{1}{l^2}C_{rr},
\end{eqnarray}
where $\eta_{AB}dx^Adx^B = -\frac{1}{l^2}dt^2 + d\phi^2$ is a fixed
metric on the cylinder and $\eta^{AB}$ is its inverse. In term of
those fields, our quantities describing the boundary gravitons are
given by:
\begin{gather}
Q=2e^\varphi, \qquad M=\frac{2}{l^2}e^\varphi \left( e^{-2\varphi}
  C_{\phi\phi} + \frac{1}{2l^2} C_{rr}\right),\\
P = -2l \dot \varphi, \qquad J = \frac{2}{l}C_{t\phi}.
\end{gather}
The lagrange multipliers take the following form:
\begin{equation}
N= \frac{r}{l} e^\varphi - \frac{l}{2}C_{tt}
e^{-\varphi}r^{-1}+o(r^{-1}), \quad N^r = O(r^{-1}), \quad N^\phi = O(r^{-2}),
\end{equation}
which leads to
\begin{equation}
f_H =\frac{Q}{2l}, \quad \tilde\chi_H = \frac{M}{2l}, \quad \psi_H = 0, \quad Y_H = 0.
\end{equation}
The associated differentiable Hamiltonian is then easily computed
\begin{equation}
	\label{eq:hamiltEBH}
H_{EBH} \approx \frac{1}{16\pi G}\oint_{\d \Sigma} d\phi \,  \frac{1}{2}
M Q.
\end{equation}
For the BTZ Black-hole \eqref{eq:BTZinf}, we have $H_{EBH} \approx m$
as expected. Using the equation of motion for $Q$, one can check that the
condition $Q>0$ is preserved under time evolution.

This set of boundary conditions possesses an asymptotic symmetry group
given by two Virasoros in semi-direct product with two current
algebras. As we have already computed the induced bracket on the boundary gravitons, we
just need to find the boundary generators in terms of $Q, P, J$ and
$M$ that are symmetry generators for the Hamiltonian $H_{EBH}$.
Let's define
\begin{gather}
\label{eq:definfcLcP}
\cL^\pm(\phi) = \frac{l}{32 \pi G} \left(\frac{1}{2} M Q \pm J\right), \qquad
\cP^\pm(\phi) = \frac{l}{32\pi G}\left( -P \pm \frac{2}{Q} \d_\phi
  Q\right), \\
\label{eq:definfcQ}
\cQ = \frac{-l}{16\pi G}\oint_{\d\Sigma} d\phi \log Q.
\end{gather}
They have the following bracket
\begin{eqnarray}
\left\{ \cL^\pm(\phi), \cL^\pm(\phi')\right\}^*  &\approx& 
\pm\cL^\pm(\phi) \,\d_\phi \delta(\phi - \phi')\mp\cL^\pm(\phi')\, \d_{\phi'} \delta(\phi' - \phi),\\
\left\{ \cL^\pm(\phi), \cP^\pm(\phi')\right\}^* &\approx & \pm\cP^\pm(\phi)
\,\d_\phi \delta(\phi - \phi') - \frac{l}{16\pi G} \, \d_\phi^2 \delta(\phi - \phi'),\\
\left\{ \cP^\pm(\phi), \cP^\pm(\phi')\right\}^* &\approx& \mp \frac{l}{16\pi G} \, \d_\phi \delta(\phi - \phi'),
\end{eqnarray}
\begin{equation}
\left\{ \cL^\pm(\phi), \cQ \right\}^* 
\approx  \frac{1}{2} \cP^\pm (\phi), \qquad\qquad
\left\{ \cP^\pm(\phi), \cQ\right\}^*  \approx -\frac{l}{32\pi G},
\end{equation}
where the rest gives zero. 
If we expand them in modes, 
\begin{equation}
\cL^\pm_m = \oint_{\d\Sigma} d\phi\, e^{\pm im \phi} \cL^\pm(\phi),
\qquad \cP^\pm_m = \oint_{\d\Sigma} d\phi\, e^{\pm im \phi} \cP^\pm(\phi),
\end{equation}
we recover the algebra obtained in 
\cite{Troessaert2013}:
\begin{equation}
\label{eq:adsextalgebra}\begin{array}{rclcrcl}
i\left\{\cL^\pm_m,\cL^\pm_n\right\}^*&=&(m-n)\cL^\pm_{m+n}, & \qquad & i\{\cL^+_m,\cL^-_n\}^*&=&0, \\
i\left\{\cL^\pm_m,\cP^\pm_n\right\}^*&=&-n \cP^\pm_{m+n} + \frac{l}{8G} im^2\delta_{m+n,0}, & \qquad &  i\{\cL^\pm_m,\cP^\mp_n\}^*&=&0, \\
i\left\{\cP^\pm_m,\cP^\pm_n\right\}^*&=&-\frac{l}{8G} m\delta_{m+n,0}, &
\qquad &  i\{\cP^+_m,\cP^-_n\}^*&=&0, \\
i\left\{\cL^\pm_m, \cQ\right\}^*& = & \frac{i}{2}\cP^\pm_m, &\qquad &
i\left\{\cP^\pm_m, \cQ \right\}^* & = & -i\frac{l}{16 G}\delta_{m,0}.
\end{array}
\end{equation}
The identification $\cP^+_0=\cP^-_0$ is also present here:
\begin{equation}
\oint_{\d \Sigma} d\phi \,\cP^+ = \oint_{\d \Sigma}d\phi\, \cP^- =
\frac{-l}{32\pi G}\oint_{\d \Sigma} d\phi \, P. 
\end{equation}

\vspace{5mm}

From this algebra, we can easily reconstruct the conserved quantities
$\cL^\pm_m(t), \cP^\pm_m(t)$ and $\cQ(t)$ where the quantities defined
in \eqref{eq:definfcLcP}-\eqref{eq:definfcQ} are their values at $t=0$. A conserved quantity $F(t)$
satisfies
\begin{equation}
\frac{\d}{\d t} F + \left\{F , H\right\}^* \approx 0,
\end{equation}
where $\frac{\d}{\d t}$ only hits the explicit dependence on
time. Using $H_{EBH} = \frac{1}{l}(\cL^+_0 + \cL^-_0)$, we
obtain:
\begin{equation}
\cL^\pm_m(t)=e^{i m \frac{t}{l}} \cL^\pm_m, \qquad \cP^\pm_m(t)=e^{i m
  \frac{t}{l}} \cP^\pm_m, \qquad \cQ(t)=\cQ + \frac{2}{l}\cP_0 t.
\end{equation}
By construction, the algebra \eqref{eq:adsextalgebra} is time
independent. 

These conserved quantities are
associated to asymptotic symmetries using the dictionary given in
\eqref{eq:asymparamkI}-\eqref{eq:asymparamkII}. For instance, the
angular momentum is
\begin{equation}
\cL^+_0 - \cL^-_0 = \frac{l}{16 \pi G} \oint_{\d \Sigma} d\phi \,J.
\end{equation}
It leads to
\begin{equation}
f=0, \qquad Y = 1, \qquad \psi = 0, \qquad \chi = 0,
\end{equation}
and then
\begin{equation}
\xi = O(r^{-3}), \qquad \xi^r = O(r^{-1}), \qquad \xi^\phi = 1 + O(r^{-2}),
\end{equation}
which is the expected rotation in $\frac{\d}{\d \phi}$ at infinity.

\subsection{Brown-Henneaux}
\label{sec:brown-henn-bound}

The original Brown-Henneaux (BH) boundary conditions are a sub-set of the
boundary conditions presented in the previous section where part of the boundary degrees of freedom are
frozen. We saw in \cite{Troessaert2013a} that such additional
boundary conditions on the phase-space can be imposed through residual
constraints on the boundary. 

The BH boundary conditions are given by 
\begin{eqnarray}
g_{rr} & = & \frac{l^2}{r^2} + C_{rr}r^{-4} + o(r^{-4}),\\
g_{rA} & = & C_{rA}r^{-3} + o(r^{-3}),\\
g_{AB} & = & r^2 \eta_{AB} + C_{AB} + o(1).
\end{eqnarray}
Our boundary variables are then easily computed. We have
\begin{gather}
Q=2, \qquad M=\frac{2}{l^2} \left(
  C_{\phi\phi} + \frac{1}{2l^2} C_{rr}\right),\\
P = 0, \qquad J = \frac{2}{l}C_{t\phi},
\end{gather}
and, for the lagrange multipliers,
\begin{equation}
N= \frac{r}{l} - \frac{l}{2}C_{tt} r^{-1}+o(r^{-1}), \quad N^r = O(r^{-1}), \quad N^\phi = O(r^{-2}).
\end{equation}
We see that the phase-space is smaller in this case: we have to impose
both $Q=2$ and $P=0$. The boundary gravitons are then completely
parametrized by the boundary fields $M$ and $J$.

 In order to describe this
phase-space, we will treat the two additional boundary conditions on the
boundary variables as constraints. This can be done by relaxing the
boundary conditions on the corresponding lagrange multipliers: we have
to relax both $\chi_H$ and $\psi_H$. Looking at the asymptotic form of
$N$, we see that $\tilde\chi_H$ is already relaxed: we have
\begin{equation}
\tilde\chi_H = \frac{1}{l} C_{tt} - \frac{1}{2l^5}C_{rr},
\end{equation}
but, this time, $C_{tt}$ is not related to $M$. Let's consider
the following relaxed asymptotics for the lagrange multipliers:
\begin{gather}
N= \frac{r}{l} - (\frac{l^2}{2}\tilde \chi_H + \frac{1}{4l^3}C_{rr})r^{-1}+o(r^{-1}), \\ N^r
= \psi_H +O(r^{-1}), \qquad N^\phi = O(r^{-2}),
\end{gather}
where both $\tilde \chi_H$ and $\psi_H$ are free to vary. The
corresponding differentiable Hamiltonian generating the additionary boundary
constraints $P=0$ and $Q=2$ is then given by:
\begin{multline}
H_{BH} =\frac{1}{16\pi G} \int_\Sigma d^2x \, \left( N \cR + N^i
  \cR_i\right) \\+ \frac{1}{16 \pi G}\oint_{\d \Sigma} d\phi \, \Big(
  M + l \psi_H P
  + l \tilde\chi_H (Q-2) \Big),
\end{multline}
The variation of the action then gives
\begin{multline}
\delta S = \int dt \int_{\Sigma}d^2x \, \left(\frac{\delta S}{\delta
    g_{ij}} \delta g_{ij} + \frac{\delta S}{\delta \pi^{ij}} \delta
  \pi^{ij} - \delta N \cR - \delta N^i \cR_i  \right)\\
 + \frac{l}{16 \pi G}\oint_{\d \Sigma} d\phi \, \Big(
  \delta\psi_H P
  + \delta\tilde\chi_H (Q-2) \Big),
\end{multline}
which is what we wanted: $\psi_H$ and $\tilde \chi_H$ are playing the
role of lagrange multipliers enforcing $Q=2$ and $P=0$.

We can now do our analysis of the boundary dynamics using the full
boundary phase-space described in section \ref{sec:dirac-brack-bound} with the
Hamiltonian:
\begin{eqnarray}
H_{BH} &\approx& \frac{1}{16 \pi G}\oint_{\d \Sigma} d\phi \, 
  M +  \frac{l}{16 \pi G}\oint_{\d \Sigma} d\phi \, \Big(\psi_H P
  +\tilde \chi_H (Q-2) \Big),\\
&\approx& H_{EBH}+  \frac{l}{16 \pi G}\oint_{\d \Sigma} d\phi \, \Big(\psi_H P
  +\tilde \chi_H (Q-2) \Big).
\end{eqnarray}
In the second line, we used the constraints $Q=2$ to recover the Hamiltonian of the
previous section \eqref{eq:hamiltEBH}. We see that the theory corresponding to the
Brown-Henneaux boundary conditions is a constrained version of the
theory associated to the boundary conditions \eqref{eq:bcconfextI}-\eqref{eq:confaddcond}.

The boundary constraints are second-class:
\begin{equation}
\left\{P(\phi), Q(\phi')-2\right\}^* \approx -\frac{16 \pi G}{l}\Big(Q(\phi)-2\Big)
\delta (\phi-\phi') - \frac{32\pi G}{l} \delta(\phi - \phi'),
\end{equation}
the other brackets being zero. It is then straightforward to compute
the induced bracket on the fully reduced phase-space. In term of $M$
and $J$, we have
\begin{eqnarray}
\left\{ M(\phi), M(\phi')
\right\}^{**} &\approx& \frac{16 \pi G}{l}\Big(J(\phi) \,\d_\phi
\delta(\phi - \phi') -J(\phi')\, \d_{\phi'} \delta(\phi' - \phi)\Big),\\
\left\{M(\phi), J(\phi') \right\}^{**} &\approx& \frac{16 \pi G}{l}\Big(M(\phi) \,\d_\phi
\delta(\phi - \phi') -M(\phi')\, \d_{\phi'} \delta(\phi' -
\phi)\Big)\nonumber\\
&& \qquad \qquad-\frac{32 \pi G}{l}\d^3_\phi \delta (\phi - \phi'),\\
\left\{J(\phi), J(\phi') \right\}^{**} &\approx& \frac{l}{16 \pi G}\Big(J(\phi) \,\d_\phi
\delta(\phi - \phi') - J(\phi')\, \d_{\phi'} \delta(\phi' - \phi)\Big),
\end{eqnarray}
where $\approx$ means in this case that we have imposed all
constraints: from both the bulk and the boundary. On this fully
reduced phase-space, the Hamiltonian is simply given by
\begin{equation}
H_{BH} \approx \frac{1}{16 \pi G}\oint_{\d \Sigma} d\phi \,  M.
\end{equation}
The two Virasoro algebras of conserved charges can be
recovered easily. Defining
\begin{equation}
\xbar \cL^\pm(\phi) = \frac{l}{32 \pi G} \Big(M(\phi) \pm J(\phi)\Big), \qquad \xbar \cL^\pm_m = \oint_{\d \Sigma}d\phi
\, e^{\pm i m \phi}\xbar\cL^\pm(\phi),
\end{equation}
we obtain the usual result
\begin{equation}
\label{eq:BHvirasoro}
i\left\{ \xbar\cL^\pm_m, \xbar\cL^\pm_n\right\}^{**} \approx (m-n) \xbar\cL^\pm_{m+n} + \frac{l}{8G}
m^3 \delta_{m+n,0}, \quad i\left\{ \xbar\cL^+_m,
  \xbar\cL^-_n\right\}^{**} \approx 0.
\end{equation}
The Virasoro generators $\xbar\cL^\pm_m$ are the generators defined on
the previous section $\cL^\pm_m$ evaluated on the constraint's surface
$Q=2$ and $P=0$. The central charge in \eqref{eq:BHvirasoro} appeared
due to the correction coming from the Dirac bracket $\{ \,,
\,\}^{**}$. The conserved charges $\cL^\pm_m(t)$ are easily computed:
\begin{equation}
\xbar\cL^\pm_m(t) = e^{im\frac{t}{l}} \xbar\cL^\pm_m.
\end{equation}

The algebra obtained here is of course just the current algebra of the
dual Liouville theory living on the boundary \cite{Coussaert1995,Henneaux2000,Rooman2001,Barnich2013}.


\subsection{Chiral}
\label{sec:chir-bound-cond}

In \cite{Compere2013}, the authors proposed a set of chiral
boundary conditions for $AdS_3$ that was extended in
\cite{Avery2014}. We will first find the Hamiltonian for the
extended version and then obtain the additional boundary constraints
corresponding to the original chiral boundary conditions. For the
extended case, the asymptotic behavior of the metric in the Fefferman-Graham
gauge can be written as 
\begin{eqnarray}
g_{rr} & = & \frac{l^2}{r^2},\\
g_{r\phi} & = & 0,\\
g_{\phi\phi} & = & r^2 (1 + F) + C_{\phi\phi} + o(1),\\
g_{rt} & = & 0,\\
g_{t\phi} & = &\frac{F}{l} r^2 + C_{t\phi} + o(1),\\
g_{tt} & = &  \frac{r^2}{l^2} (-1 + F) + \Delta- l^{-2} C_{\phi\phi} + 2 l^{-1} C_{t\phi} + o(1),
\end{eqnarray}
where $F$ is a function of $t$ and $\phi$ and $\Delta$ is a fixed constant. As
we assumed $\bar \gamma>0$, this
means that we are studying the case $F>-1$ only. A straighforward
computation leads to the following values for our quantities
describing the boundary gravitons:
\begin{gather}
Q = 2 \sqrt{1+F}, \qquad M= \frac{2}{l^2}\frac{C_{\phi\phi}}{\sqrt{1+F}}+\frac{1}{l^4}C_{rr}\sqrt{1+F}, \\
P = -l \d_t F + \frac{2+F}{1+F}\d_\phi F, \qquad
J=\frac{2}{l}C_{t\phi}(1+F) - \frac{2}{l^2}C_{\phi\phi}F,
\end{gather}
associated to the lagrange multipliers:
\begin{gather}
N = \frac{r}{l}\frac{1}{\sqrt{1+F}} +
\frac{l}{2r}\sqrt{1+F}\left(-\Delta
  +\frac{C_{\phi\phi}}{l^2}\frac{1+2F}{(1+F)^2}
  -\frac{2}{l}\frac{C_{t\phi}}{1+F}\right)+o(r^{-1}),\\
N^\phi = \frac{1}{l}\frac{F}{(1+F)}+O(r^{-2}), \qquad N^r = O(r^{-1}).
\end{gather}
This leads to
\begin{gather}
f_H=\frac{2}{l} \frac{1}{Q}, \qquad Y_H=\frac{1}{l} - \frac{4}{l} \frac{1}{Q^2},
\qquad \Psi_H=0,\\
\chi_H = \frac{\Delta}{2l} Q + \frac{8}{l}\frac{J}{Q^3} - \frac{2}{l}
\frac{M}{Q^2},
\end{gather}
and to the Hamiltonian:
\begin{equation}
\label{eq:HamilChiral}
H_{EC} \approx \frac{1}{16 \pi G}\oint_{\d \Sigma} d\phi \, \left(J -
	4\frac{J}{Q^2} +
2 \frac{M}{Q} +\Delta \frac{Q^2}{4}\right).
\end{equation}

The equation of motion for $F$ is given by
\begin{equation}
	\Delta\frac{\d }{\d x^-} F+ \left(\frac{\d}{\d x^-}\right)^3F = 0.
\end{equation}
where $x^- = \frac{t}{l} - \phi$. In general, we cannot expect the time
evolution to preserve the condition $F>-1$. The breaking of this condition
means that the surfaces of constant $t$ are not space-like and
our ADM split is not valid anymore. However, if the initial conditions
satisfy $F>-1$ then it will stay valid close to $t_0$ and, in this
neighborhood, we can still apply our analysis. 

\vspace{5mm}

In \cite{Avery2014}, the authors showed that, for $\Delta<0$,
the algebra of the charges is given by the semi-direct product of a
Virasoro algebra with a $sl(2,\RR)$ current algebra. Functionals
of the boundary gravitons reproducing this result are built from
\begin{gather}
L_C(\phi) =\frac{l}{16\pi G}J - \d_\phi\cP^+, \quad
T^0_C(\phi)  =  \cP^+,\\
T^+_C(\phi) = \frac{l}{4\pi G}\left(\frac{M}{Q}- 2 \frac{J}{Q^2}\right), \quad
T^-_C(\phi) =\frac{l}{16\pi G}\frac{-Q^2}{8},
\end{gather}
where $\cP^+(\phi)$ was defined in \eqref{eq:definfcLcP}. The brackets
of these new quantities are given by:
\begin{eqnarray}
\left\{L_C(\phi), L_C(\phi') \right\}^* & \approx & L_C(\phi)
\d_\phi \delta(\phi - \phi') - L_C(\phi')
\d_\phi' \delta(\phi' - \phi) \nonumber \\ && \qquad- \frac{l}{16 \pi G} \d_\phi^3
\delta(\phi - \phi'),\\
\left\{L_C(\phi), T_C^a(\phi') \right\}^* & \approx & T_C^a(\phi)
\d_\phi \delta(\phi - \phi'),\\
\left\{T^a_C(\phi), T^b_C(\phi') \right\}^* & \approx &f^{ab}_c
T_C^c(\phi) \delta (\phi-\phi') +\frac{l }{16\pi G}\eta^{ab} \d_\phi
\delta (\phi-\phi'),
\end{eqnarray}
where $a,b,c=+,-,0$. The current algebra is characterized by
\begin{equation}
f^{0+}_+=-1, \quad f^{0-}_-=1, \quad f^{+-}_0=2, \quad
\eta^{00}=-1, \quad \eta^{+-}=2,
\end{equation}
with all the other components equal to zero.
If we develop in modes:
\begin{gather}
L_{m} = \oint_{\d\Sigma}d\phi \, e^{im\phi} L_C(\phi),\quad T^a_{m} =
\oint_{\d\Sigma}d\phi \, e^{im\phi} T^a_C(\phi),
\end{gather}
we recover the algebra of the charges found in \cite{Avery2014}
\begin{eqnarray}
\label{eq:ChiralChargeI}
i \left\{ L_{m}, L_{n}\right\} & = & (m-n)L_{m+n} +\frac{l}{8 G}m^3
\delta_{m+n, 0},\\
i \left\{ L_{m}, T^a_{n}\right\} & = & -n  T^a_{m+n},\\
\label{eq:CHiralChargeIII}
i \left\{  T^a_{m}, T^b_{n}\right\} & = & i f^{ab}_c
 T^c_{m+n}+\frac{l }{8 G}\eta^{ab} m \delta_{m+n,0}.
\end{eqnarray}

In terms of these generators, the Hamiltonian is given by,
\begin{equation}
	H =\frac{1}{l}(L_0 + \frac{1}{2} T^+_0 - 2\Delta T^-_0).
\end{equation}
The conserved quantities are easily built by adding an explicit time
dependence to each mode. If $\Delta =- \alpha^2$, we get
\begin{eqnarray}
	\tilde L_m(t) &=& L_m e^{im\frac{t}{l}}, \\ 
	\tilde T^0_m(t) &=&T^0_m e^{im\frac{t}{l}} \cos(2\alpha \frac{t}{l}) + 
	\frac{1}{4\alpha}T^+_m e^{im\frac{t}{l}} \sin(2\alpha \frac{t}{l})\nonumber \\ &&
	 \qquad-\alpha T^-_m e^{im\frac{t}{l}} \sin(2\alpha \frac{t}{l}),\\
	\tilde T^+_m(t) &=& T^+_m e^{im\frac{t}{l}}\cos^2(\alpha \frac{t}{l})+
	4\alpha^2 T^-_m e^{im\frac{t}{l}}\sin^2(\alpha \frac{t}{l})\nonumber\\ && \qquad-
	2\alpha T^0_m e^{im\frac{t}{l}} \sin(2\alpha \frac{t}{l}),\\
	\tilde T^-_m(t) &=& \frac{1}{4\alpha^2}T^+_m e^{im\frac{t}{l}}\sin^2(\alpha \frac{t}{l}) +
	T^-_me^{im\frac{t}{l}}\cos^2(\alpha \frac{t}{l})\nonumber
	\\ && \qquad + \frac{1}{2\alpha} T^0_m e^{im\frac{t}{l}} \sin(2\alpha \frac{t}{l}).
\end{eqnarray}
The cases $\Delta=0$ and $\Delta <0$ can be obtained in a similar way.

The charges we obtained here are not the one obtained in
\cite{Avery2014}. However, we built these because the are well adapted to the
constraints analysis that we will do in the next section.

\subsection{Constrained Chiral}
\label{sec:constr-chir-bound}

The original chiral boundary conditions introduced in
\cite{Compere2013} are a subset of the one introduced in the
previous section with the additional condition
\begin{equation}
\d_t F - \frac{1}{l} \d_\phi F = 0.
\end{equation}
A good point here is that this extra condition garanties the preservation of
$F>-1$ under time evolution. 
This can easily be rewritten as
\begin{equation}
\label{eq:chiralconstraint}
T^0_C=\cP^+ = \frac{l}{32 \pi G} \left(-P + \frac{2}{Q}\d_\phi
Q\right) = 0.
\end{equation}
The boundary theory associated to these restricted boundary conditions can be
described by 
a theory built from the Hamiltonian
\eqref{eq:HamilChiral} with the constraint $T^0_C\approx 0$.

For simplicity, the rest of the analysis will be done using Fourier modes. The
primary constraints are $T^0_m\approx 0$ for all $m$. They lead to secondary
constraints:
\begin{gather}
	\left\{ T^0_{m}, lH\right\} =  -imT^0_m
	-\frac{1}{2}T^+_m-2\Delta T^-_m,\\
	\Rightarrow K^+_m \equiv \frac{1}{2}T^+_m+2\Delta T^-_m\approx 0.
\end{gather}
With these extra constraints, the set is complete:
\begin{equation}
	\left\{K^+_m, lH \right\} = -imK^+_m -4\Delta T^0_m \approx 0.
\end{equation}
Their algebra is complicated and it is difficult to form the Dirac bracket. However, when $\Delta \ne
0$, $L_m$ and $K^-_m\equiv \frac{1}{2}T^+_m-2\Delta T^-_m$
form a complete set
of gauge-invariant quantities.
The reduced phase-space is parametrized by $L_m$ and $K^-_m$ with
\begin{eqnarray}
	i\left\{L_m, L_n \right\} &=& (m-n)L_{m+n} + \frac{l}{8G}
	m^3\delta_{m+n,0},\\
	i\left\{L_m, K^-_n \right\} &=& -n K^-_{m+n},\\
	i\left\{K^-_m, K^-_n \right\} &=& -\frac{\Delta l}{2G} m\delta_{m+n,0},
\end{eqnarray}
and the Hamiltonian given by
\begin{equation}
	H = \frac{1}{l}(L_0 + K^-_0).
\end{equation}

To make the link with the charges obtained in equations (2.15) and (2.16) of 
\cite{Compere2013}, we define
\begin{equation}
	\hat L_m=L_m + \frac{1}{2}K_m^--\frac{l\Delta}{16G}\delta_{m,0},\qquad \hat
	P_m=\frac{1}{2}K_m^--\frac{l\Delta}{16G}\delta_{m,0}.
\end{equation}
which leads to the following algebra
\begin{eqnarray}
	i\left\{\hat L_m, \hat L_n \right\} &=& (m-n)\hat L_{m+n} + \frac{l}{8G}
	m^3\delta_{m+n,0},\\
	\label{eq:extchiralCompII}
	i\left\{\hat L_m, \hat P_n \right\} &=& -n \hat
	P_{m+n}
			-\frac{\Delta l}{16G} m\delta_{m+n,0},\\
	i\left\{\hat P_m, \hat P_n\right\} &=& -\frac{\Delta l}{8G} m\delta_{m+n,0}.
\end{eqnarray}
The algebra we obtain here is the algebra of the generators
written in equations (2.15) and (2.16) of 
\cite{Compere2013}. However, the difference between the extension obtained
here and the one written in equations (2.17)-(2.19) of \cite{Compere2013} is a
redefiniton of the zero mode $\hat P_0 \rightarrow \hat P_0
-\frac{l\Delta}{16G}\delta_{m,0}$.
This change of basis absorbs the extension in \eqref{eq:extchiralCompII} and brings back the algebra to the canonical form
with the following central charge and level 
\begin{equation}
	c_R=\frac{3l}{2G}, \qquad k_{KM}=-\frac{l\Delta}{4G}.
\end{equation}

\section*{Acknowledgements}
\label{sec:acknowledgements}

I would like to thank G. Barnich, H. Gonz\'alez, P. Ritter and D. Tempo for
useful discussions. This work is founded by the fundecyt postdoctoral grant
3140125. The Centro de Estudios Cient\'ificos (CECs) is funded by the Chilean
Government through the Centers of Excellence Base Financing Program of
Conicyt.

\newpage
\appendix
\section{Radial decomposition}
\label{sec:ADMspa}

Let's assume that we have spatial coordinates given by $x^i=(r,\phi)$
. We introduce:
\begin{equation}
\gamma \equiv g_{\phi\phi}, \quad \lambda_\phi \equiv g_{r\phi}, \quad
\lambda^\phi \equiv \lambda_\phi \gamma^{-1}, \quad \lambda \equiv \frac{1}{\sqrt{g^{rr}}}.
\end{equation}
The metric and its inverse take the form:
\begin{equation}
g_{ij}= \left(\begin{array}{cc}
\lambda^2 + \lambda^\phi\lambda_\phi & \lambda_\phi \\
\lambda_\phi & \gamma
  \end{array}
\right),\quad 
g^{ij}= \left(\begin{array}{cc}
\frac{1}{\lambda^2} & -\frac{\lambda^\phi}{\lambda^2} \\
-\frac{\lambda^\phi}{\lambda^2} & \gamma^{-1}+ \frac{\lambda^\phi\lambda^\phi}{\lambda^2}
  \end{array}
\right),
\end{equation}
where we used $\gamma$ and its inverse $\gamma^{-1}$ to raise and
lower the angular indices $\phi$

Introducing the extrinsic curvature of the (1)-spheres $K_{\phi\phi}$, we can write
all the Christoffel symbols:
\begin{eqnarray}
K_{\phi\phi} & =  & \frac{1}{2 \lambda} \left( - \d_r \gamma + 2D_\phi
\lambda_\phi\right)\\
K = \gamma^{-1} K_{\phi\phi} &=& \frac{1}{2\lambda} \left((-\d_r
+\lambda^\phi\d_\phi)\log \gamma + 2\d_\phi \lambda^\phi\right),\\
\Gamma^r_{\phi\phi} & = & \frac{1}{\lambda}\gamma K \\
\Gamma^\phi_{\phi\phi} & = & \frac{1}{2}\d_\phi \log \gamma -
\frac{\lambda_\phi}{\lambda} K\\
\Gamma^r_{r\phi} & = & \frac{1}{\lambda} \left( \d_\phi \lambda + K
\lambda_\phi\right) \\
\Gamma^r_{rr} & = &\frac{1}{\lambda} \d_r \lambda
+\frac{\lambda^\phi}{\lambda} \left( \d_\phi \lambda + K
  \lambda_\phi\right) \\ 
\Gamma^\phi_{r\phi} & = & -\frac{\lambda^\phi}{\lambda} \left( \d_\phi \lambda
  +K
  \lambda_\phi\right) + D_\phi \lambda^\phi - \lambda K \\
\Gamma^\phi_{rr} & = & -\lambda \left( \gamma^{-1} + \frac{\lambda^\phi
    \lambda^\phi}{\lambda^2}\right) \left( \d_\phi\lambda +
 K\lambda_\phi\right) \nonumber \\ && \qquad- \lambda^\phi \left( D_\phi \lambda^\phi - \lambda
  K\right) - \frac{\lambda^\phi}{\lambda} \d_r \lambda +
	 \gamma^{-1}
\d_r \lambda_\phi
\end{eqnarray}
where $D_\phi$ is the covariant derivative associated to $\gamma$.

\vspace{3mm}

We have introduced new variables to describe $g_{ij}$. Some computations can
be simplified introducing their canonical conjugates:
\begin{equation}
\theta \equiv 2 \lambda \pi^{rr}, \quad \theta_\phi \equiv 2 \gamma \pi^{r\phi} +
2 \lambda_\phi \pi^{rr}, \quad \sigma \equiv \pi^{\phi\phi} + 2
\lambda^\phi
\pi^{\phi r} + \lambda^\phi \lambda^\phi \pi^{rr}.
\end{equation}
They satisfy:
\begin{equation}
\pi^{ij}\delta g_{ij} = \theta \delta \lambda + \theta_\phi \delta \lambda^\phi +
\sigma \delta \gamma,
\end{equation}
\begin{equation}
\left\{ \lambda(x), \theta(x')\right\} = \delta^2(x-x'), \quad
\left\{ \lambda^\phi(x), \theta_\phi(x')\right\} = \delta^2(x-x'),
\end{equation}
\begin{equation}
\left\{ \gamma(x), \sigma(x')\right\} = \delta^2(x-x'),
\end{equation}
all the other Poisson brackets being zero.

\section{Gauge transformations}
\label{sec:gaugetransf}

It is useful to have the gauge transformation laws of the new dynamical fields
defined in appendix \ref{sec:ADMspa}. We
will only consider here transformations with parameters $\xi, \xi^i$
independent of the canonical variables.
Straightforward but long computations lead to
\begin{itemize}
\item for the metric, we obtain:
\begin{eqnarray}
	\delta_\xi \lambda &=& -\xi \sqrt \gamma \sigma + \d_r (\lambda \xi^r) + \xi^\phi \d_\phi \lambda -
 \lambda \lambda^\phi \d_\phi \xi^r,\\
\delta \lambda^{\phi} &=& \lambda \xi \gamma^{-\frac{3}{2}} \theta_\phi 
+ \d_r (\xi^\phi + \xi^r \lambda^\phi) \nonumber \\ && \quad- \lambda^\phi \d_\phi
\xi^\phi + \xi^\phi \d_\phi \lambda^\phi+ \d_\phi \xi^r \left( \lambda^2 \gamma^{-1} -
  (\lambda^\phi)^2\right),\\
  \delta_\xi \gamma &=& - \xi \sqrt{\gamma}\theta + \xi^i \d_i \gamma + 2
\d_\phi \xi^\phi \gamma + 2 \d_\phi \xi^r \lambda_\phi,
\end{eqnarray}
\item for the trace of the extrinsic curvature $K$
\begin{eqnarray}
	\delta_\xi K& = &\frac{\xi}{\lambda} \sqrt \gamma \sigma K - \frac{1}{2\lambda}
	(\lambda^\phi \d_\phi - \d_r) \left( \frac{\theta\xi}{\sqrt \gamma}\right)
	+ \frac{1}{\lambda} D_\phi \left(\lambda\xi
	\frac{\theta^\phi}{\sqrt\gamma}\right)\nonumber\\
	&& +\xi^r \d_r K + 2 \d_\phi \lambda \gamma^{-1} \d_\phi \xi^r +
	\lambda D_\phi D^\phi \xi^r + \xi^\phi \d_\phi K.
	\label{eq:transfoK}
\end{eqnarray}
\item the variations of the momenta are given by
\begin{eqnarray}
\delta_\xi \theta & = & -2 \sqrt{\gamma}\xi \Lambda -
	2\sqrt{\gamma} D_\phi D^\phi \xi + \frac{2}{\lambda} \sqrt \gamma
	K(\d_r-\lambda^\phi\d_\phi) \xi - \frac{\xi}{2}
	\sqrt{\gamma} (\gamma^{-1}\theta_\phi)^2 \nonumber \\ &&
\quad + \xi^r \d_r \theta + \d_\phi (\xi^\phi \theta) + \lambda^\phi \d_\phi
\xi^r \theta - 2\lambda \d_\phi \xi^r \gamma^{-1}\theta_\phi,\\
\delta_\xi \theta_\phi &=&2 \sqrt \gamma \d_\phi\left[\frac{1}{\lambda} (\d_r -
\lambda^\phi\d_\phi) \xi\right]+ 2 \sqrt \gamma K \d_\phi \xi \nonumber \\ &&  +
\xi^r \d_r \theta_\phi + \d_\phi (\xi^\phi \theta_\phi) + \d_\phi \xi^\phi
\theta_\phi  + 2 \d_\phi \xi^r \left( \frac{\lambda \theta}{2} -
  \gamma\sigma + \lambda^\phi \theta_\phi \right),\\
  \delta_\xi \sigma & = & \d_i(\xi^i \sigma) - 2 \sigma (\d_\phi \xi^\phi +
  \lambda^\phi \d_\phi \xi^r) +\lambda^2 \gamma^{-2} \d_\phi \xi^r \theta_\phi
   -  \frac{\xi}{\sqrt \gamma}\lambda \Lambda \nonumber \\ 
  &&+\frac{1}{2} \frac{\xi}{\sqrt\gamma} \sigma \theta
   + \frac{3}{4} \frac{\xi}{\sqrt\gamma}\lambda (\gamma^{-1} \theta_\phi)^2
   - \frac{1}{\sqrt\gamma}\d_\phi\lambda \gamma^{-1}\d_\phi \xi \nonumber \\ 
  &&-\frac{1}{\sqrt \gamma} (\d_r- \lambda^\phi\d_\phi)\left[\frac{1}{\lambda} (\d_r
  - \lambda^\phi \d_\phi) \xi\right] ,
\end{eqnarray}
\end{itemize}

\vspace{5mm}

In section \ref{sec:inf-diff-gauge-gener}, we showed that the gauge transformations preserving the asymptotics are given by
\begin{eqnarray}
\label{eq:allowedtest}
\xi & = & r f + \kappa + O(r^{-3}),\\
\xi^r & = & r \psi + O(r^{-1}),\\
\xi^\phi & = & Y + O(r^{-3}),
\end{eqnarray}
where
\begin{eqnarray}
	\kappa &=& - \frac{l^2}{2r} \chi - r\int_r^\infty dr' j(r')=O(r^{-1}), \\
j &=& \frac{\tilde \lambda}{l} f + \lambda^\phi\d_\phi f + \frac{l}{r}( K+
  \frac{1}{l}) f = O(r^{-3}).
\end{eqnarray}
The
four functions $f, \chi, \psi$ and $Y$ are independent of $r$. A useful result
is:
\begin{equation}
	\frac{1}{r}(\d_r - \frac{1}{r}) \kappa = \frac{l^2}{r^3} \chi + \frac{\tilde
	\lambda}{l} f + \lambda^\phi \d_\phi f + \frac{l}{r} (K + \frac{1}{l})
	f
\end{equation}

Let's assume that our improper gauge parameters are independent of the
dynamical variables. The bracket of two gauge transformations is
\begin{equation}
	\left[\xi_1, \xi_2\right]^\mu \approx \left[\xi_1, \xi_2\right]^\mu_{SD} -
	\delta^z_1\xi^\mu_2 + \delta^z_2\xi^\mu_1.
\end{equation}
where the variation $\delta^z$ hits the dynamical variables only and $\xi^\mu =
(\xi, \xi^r, \xi^\phi)$.
Long computations lead to
\begin{eqnarray}
	\hat f & = & Y_1\d_\phi f_2 +\psi_1f_2 - (1\leftrightarrow 2)\\
	\hat \chi & = & Y_1\d_\phi\chi_2 - \psi_1 \chi_2 - D_\phi(
	f_2\bar\gamma^{-1}\d_\phi\psi_1) - \frac{P}{2\sqrt{\bar\gamma}}f_1
	\chi_2 -  (1\leftrightarrow 2)\\
	\hat Y & = & Y_1\d_\phi Y_2 + \bar\gamma^{-1}f_1 \d_\phi f_2 
	-  (1\leftrightarrow 2)\\
	\hat \psi & = & Y_1\d_\phi \psi_2 + f_1\chi_2 - (1\leftrightarrow 2),
\end{eqnarray}
where the hatted quantities parametrize the resulting transformation $[\xi_1,
\xi_2]$.

\section{Fefferman-Graham gauge fixing}
\label{sec:FGfixing}

Some computations are a lot easier when the gauge is fixed. In this work, we
are using the Fefferman-Graham choice
\cite{Fefferman1985,Graham1991,Skenderis2000,Graham1999,Rooman1999,Bautier2000}. However, due to the relaxed asymptotics
and the fact that we are working in the hamiltonian framework, it is
not clear if this gauge can be reached with a proper gauge transformation.

In coordinates $x^\mu = (r, t, \phi)$ the FG gauge is given by
$g_{rr}=\frac{l^2}{r^2}$ and $g_{rt}=0=g_{r\phi}$. On the canonical variables,
these conditions become
\begin{equation}
	g_{rr} = \frac{l^2}{r^2}, \quad g_{r\phi} = 0, \quad \pi^{\phi\phi} =
	0.
\end{equation}
The question now is how can we send a configuration satifying our relaxed
boundary conditions \eqref{eq:assympg} and \eqref{eq:assympp} onto this gauge-fixed surface?

\vspace{5mm}

Let's introduce auxiliary quantities $\tilde N$, $\tilde N^i$ satisfying
\begin{equation}
	\tilde N = \frac{r}{l} + O(r^{-3}), \quad \tilde N^r = O(r^{-1}),
	\quad \tilde N^\phi = O(r^{-2}).
\end{equation}
The exact value of these fields is not important. Using these to generate a
time evolution along $s$, we can build an
auxiliary 3 dimensional metric with coordinates $x^\mu = (s, r, \phi)$:
\begin{gather}
	\tilde g_{ss} = -\tilde N^2 + \tilde N^i g_{ij} \tilde N^j, \quad \tilde
	g_{si} = g_{ij} \tilde N^j,\quad \tilde g_{ij} = g_{ij},\\
	\d_s g_{ij} = \left\{g_{ij}, \int (\tilde N \cR + \tilde N_i\cR_i
	\right\},\\
	\d_s \pi^{ij} = \left\{\pi^{ij}, \int (\tilde N \cR + \tilde N_i\cR_i
	\right\}.
\end{gather}
Because the lagrange multipliers we chose preserve the asymptotic behavior of
the canonical fields under "time" evolution, the auxiliary metric takes the form
\begin{gather}
	\tilde g_{rr} = \frac{l^2}{r^2} + O(r^{-4}), \quad \tilde g_{rs} =
	O(r^{-3}), \quad \tilde g_{r\phi} = O(r^{-1}),\\
	\tilde g_{ss} = -\frac{r^2}{l^2} + O(r^{-2}), \quad \tilde g_{s\phi} =
	O(1), \quad \tilde g_{\phi\phi} = r^2 \bar \gamma + O(1).
\end{gather}
This can be put into the Fefferman-Graham form using a change of
coordinates of the form
\begin{eqnarray}
	r & = & r' + O(r'^{-1}),\\
	s & = & s' + O(r'^{-4}),\\
	\phi & = & \phi' + O(r'^{-2}).
\end{eqnarray}
This transformation is the exponential of the transformation generated by
a vector with the following asymptotic behavior
\begin{equation}
	{}^{(3)}\xi^r = O(r^{-1}), \quad {}^{(3)}\xi^s = O(r^{-4}), \quad
	{}^{(3)}\xi^\phi = O(r^{-2}),
\end{equation}
which, brought back to the hamiltonian formalism using the lagrange
multipliers $\tilde N$ and $\tilde N^i$, is a proper gauge transformation:
\begin{equation}
	\xi = O(r^{-3}), \quad \xi^r = O(r^{-1}), \quad \xi^\phi = O(r^{-2}).
\end{equation}

\bibliography{../../../Docear/_data/14C4C119A9EB7LK4NIEX6AWNRSXHJH2EOCOG/default_files/Physics}

\end{document}